\documentclass[a4paper,11pt]{article}
\usepackage{pos}

\newcommand{\mincas}{{\tt MINCAS}}
\newcommand{\katie}{{\sf KATIE}}

\title{In-medium transverse momentum
broadening effects on di-jet
observables
}

\author*[a]{Martin Rohrmoser}
\author[a]{Krzysztof Kutak}
\author[a]{Andreas van Hameren}
\author[b]{Wies\l{}aw P\l{}aczek}
\author{Konrad Tywoniuk}

\affiliation[a]{Institute of Nuclear Physics,\\
 Polish Academy of Sciences,ul. Radzikowskiego 152, 31-342 Krak\'ow, Poland}

\affiliation[b]{Institute of Applied Computer Science, Jagiellonian University,\\
ul. \L{}ojasiewicza 11, 30-348 Krak\'ow, Poland}

\affiliation[c]{Department of Physics and Technology,\\
 University of Bergen, 5007 Bergen, Norway}

\emailAdd{rohrmoser.martin1987@gmail.com}
\emailAdd{krzysztof.kutak@ifj.edu.pl}
\emailAdd{Andre.Hameren@ifj.edu.pl}
\emailAdd{Wieslaw.Placzek@uj.edu.pl}
\emailAdd{Konrad.Tywoniuk@uib.no}

\abstract{Heavy ion collisions at high energies can be used as an interesting way to recreate and study the medium of the quark-gluon plasma (QGP). We particularly investigate the jets produced in hard binary collisions and their interactions with a tentative medium. These jets were obtained numerically from the Monte-Carlo simulations of hard collisions using the \katie -algorithm~\cite{vanHameren:2016kkz}, where parton momenta within the colliding nucleons were describe by means of unintegrated parton distribution functions (uPDF). We evolved these jets within a medium that contains both, transverse kicks (yielding a broadening in momentum transvers to the jet-axis) as well as medium induced radiation within the \mincas -algorithm~\cite{Kutak:2018dim} following the works of~\cite{Blaizot:2012fh,Blaizot:2013vha}. We produce qualitative results for the decorrelation of dijets. In particular, we study deviations from a transverse momentum broadening that follows a Gaussian distribution.}

\FullConference{%
  HardProbes2020\\
  1-6 June 2020\\
  Austin, Texas}

\begin{document}
\maketitle

%
We study jet-pairs produced in hard partonic collisions during the early stages of ultrarelativistic heavy ion collisions. 
%
In order to study the ultrarelativistic nuclear collisions, we first describe di-jet production in $p$-$p$ collisions, where we assume that a QGP-medium is not formed, and later on extend 
this description to include medium effects.  
In our study of di-jet observables we focus on distributions of jet-pairs in the transverse plane, orthogonal to the collision axis of the incoming nucleons.
In this case the momentum components of the incoming hard partons transverse to the collision axis cannot be neglected, since they contribute considerably to the deviations of pure 
back-to-back emission of the di-jet pairs.
Thus, we describe the di-jet production cross section $\sigma_{hard}$ using the following factorization formula
\begin{eqnarray}\label{LO_kt-factorisation} 
\frac{d \sigma_{hard}}{d y_1 d y_2 d^2q_{1T} d^2q_{2T}} &=&
\int \frac{d^2 k_{1T}}{\pi} \frac{d^2 k_{2T}}{\pi}
\frac{1}{16 \pi^2 (x_1 x_2 s)^2} \; \overline{ | {\cal M}^{\mathrm{off-shell}}_{g^* g^* \to g g} |^2}
 \\  
&& \times  \; \delta^{2} \left( \vec{k}_{1T} + \vec{k}_{2T} 
                 - \vec{q}_{1T} - \vec{q}_{2T} \right) \;
{\cal F}_g(x_1,k_{1T}^2,\mu_{F}^2) \; {\cal F}_g(x_2,k_{2T}^2,\mu_{F}^2) \; \nonumber ,   
\end{eqnarray}
where ${\cal F}_g(x_i,k_{iT}^2,\mu_{F}^2)$ (for $i=1$, $2$) are unintegrated parton distribution functions at factorization scale $\mu_F$, which give the distribution of the transverse momenta $k_{iT}$ and the momentum fraction $x_i$ in direction of the collision axis.
The momentum fractions $x_i$, the rapidities $y_i$ and the transverse momenta $q_{iT}$ of the outgoing particles are related to one another as
$
x_1 = \frac{q_{1T}}{\sqrt{s}}\exp( y_1) 
     + \frac{q_{2T}}{\sqrt{s}}\exp( y_2)$, and 
$x_2 = \frac{q_{1T}}{\sqrt{s}}\exp(-y_1) 
     + \frac{q_{2T}}{\sqrt{s}}\exp(-y_2)\,. \nonumber
$
${\cal M}^{\mathrm{off-shell}}_{g^* g^* \to g \bar g}$ is the matrix element for the collision of virtual incoming particles and outgoing particles that are on the mass shell.
We obtained our numerical results for $p$-$p$ collisions via an implementation of Eq.~(\ref{LO_kt-factorisation}) in the Monte-Carlo program \katie.

For the description of di-jet production in heavy ion collisions, it can be approximated that the probability densities for di-jet production in heavy-ion collisions factorizes into a density for the initial hard scattering and a density for the jet-medium interactions, since either of these subprocesses involves largely different scales of momentum transfer.
We again describe the hard scattering using the factorization formula Eq.~(\ref{LO_kt-factorisation}).
We consider an in-medium jet-evolution for which, under the assumption that transverse momentum transfers from the medium to the jet particles are small, the following evolution equation was found in~\cite{Blaizot:2014rla} for distributions of the leading jet particle
\begin{equation}
\begin{aligned}
\frac{\partial}{\partial t} D(\tilde{x},\mathbf{l},t) = & \: \frac{1}{t^*} \int_0^1 dz\, {\cal K}(z) \left[\frac{1}{z^2}\sqrt{\frac{z}{\tilde{x}}}\, D\left(\frac{\tilde{x}}{z},\frac{\mathbf{l}}{z},t\right)\theta(z-\tilde{x}) 
- \frac{z}{\sqrt{\tilde{x}}}\, D(\tilde{x},\mathbf{l},t) \right] \\
+& \int \frac{d^2\mathbf{q}}{(2\pi)^2} \,C(\mathbf{q})\, D(\tilde{x},\mathbf{l}-\mathbf{q},t),
\end{aligned}
\label{eq:ktee1}
\end{equation}
with 
$
 \frac{1}{t^*}  =  \frac{\alpha_s N_c}{\pi}\sqrt{\frac{\hat{q}}{E}}$
defined via the quenching parameter $\hat{q}$ and the energy of the incoming parton $E$. 
%
%
$D(\tilde{x},\mathbf{l},t)$ is defined  as an energy density with $\tilde{x}$ the fraction of energy $E$ that the jet-particle retains  and $\mathbf{l}$  its momentum component orthogonal to the momentum of the incoming particle. 
%
%
The splitting kernel takes the form
\begin{equation}
{\cal K}(z) = \frac{\left[f(z)\right]^{5/2}}{\left[z(1-z)\right]^{3/2}}, 
\quad   f(z) = 1 - z + z^2, 
\qquad  0 \leq z \leq 1, 
\label{eq:kernel1}
\end{equation}
The scattering kernel is 
\begin{equation}
C(\mathbf{q}) = w(\mathbf{q}) - \delta(\mathbf{q}) \int d^2\mathbf{q'}\,w(\mathbf{q'})\,.
\label{eq:Cq}
\end{equation}
%
We consider here a scattering off medium particles of a form~\cite{Aurenche:2002pd} such that 
\begin{equation}
 w(\mathbf{q}) = \frac{16\pi^2\alpha_s^2N_cn}{\mathbf{q}^2(\mathbf{q}^2+m_D^2)}\,,
\label{eq:wq2}
\end{equation}
with $m_D$ the medium quasi-particle Debye mass.
After integration of Eq.~(\ref{eq:ktee1}) over the transverse momenta, one obtains 
\begin{equation}
\begin{aligned}
\frac{\partial}{\partial t} D(\tilde{x},t) = & \: \frac{1}{t^*} \int_0^1 dz\, {\cal K}(z) \left[\sqrt{\frac{z}{\tilde{x}}}\, D\left(\frac{\tilde{x}}{z},t\right)\theta(z-\tilde{x}) 
- \frac{z}{\sqrt{\tilde{x}}}\, D(\tilde{x},t) \right] \,,
\end{aligned}
\label{eq:qee1}
\end{equation}
It was shown in~\cite{Kutak:2018dim} that both Eqs.~(\ref{eq:ktee1}) and~(\ref{eq:qee1}) can be written as integral equations, which can be solved numerically via a Monte-Carlo algorithm, as it was done by the \mincas -program~\cite{Kutak:2018dim}. 

In our approach~\cite{vanHameren:2019xtr} we obtained the four momenta of the gluons produced in the hard collisions via the \katie -program and then the changes in gluon momenta due to in-medium evolution via the \mincas -program.
We parametrize the medium as follows: $\hat{q}=0.29$~GeV$^2$/fm, $n=0.08$~GeV$^3$, $m_D=0.61$~GeV
. These parameters are estimates for a medium of constant temperature $T=250$~MeV (cf.~\cite{vanHameren:2019xtr} for further explanations). The particles evolve over a time of $t_L=5$~fm/c in the medium.

We obtained numerical results for the azimuthal angular decorrelations $\frac{dN}{d\Delta\phi}$ ($N$ is the number of di-jets; $\Delta \phi$ is the difference of the azimuthal angles of the two outgoing jet-momenta). 
We compared the following three cases
\begin{enumerate}
\item A case without jet-medium interactions, referred to as the "vacuum case", where the outgoing di-jet momenta where obtained via the \katie -algorithm
\item A case with jet-medium interactions, where the in-medium jet-evolution follows Eq.~(\ref{eq:ktee1}) . We refer to this case as "non-Gaussian $k_T$ broadening".
\item 
A case referred to as "Gaussian $k_T$ broadening", where the distribution of momentum fractions $\tilde{x}$ follow Eq.~(\ref{eq:qee1}) ,
while the transverse momentum components $\mathbf{l}$ are selected from a Gaussian distribution $P(||\mathbf{l}||)$, i.e.:
  $ 
        P(||\mathbf{l}||)=\frac{1}{\sqrt{2\pi\hat{q}t_L}}\;
        \exp\left(-\frac{\mathbf{l}^2}{2\hat{q}t_L}\right).
$
\end{enumerate}

\begin{figure}
\includegraphics[scale=0.4,clip=true,trim=20 5 45 0]{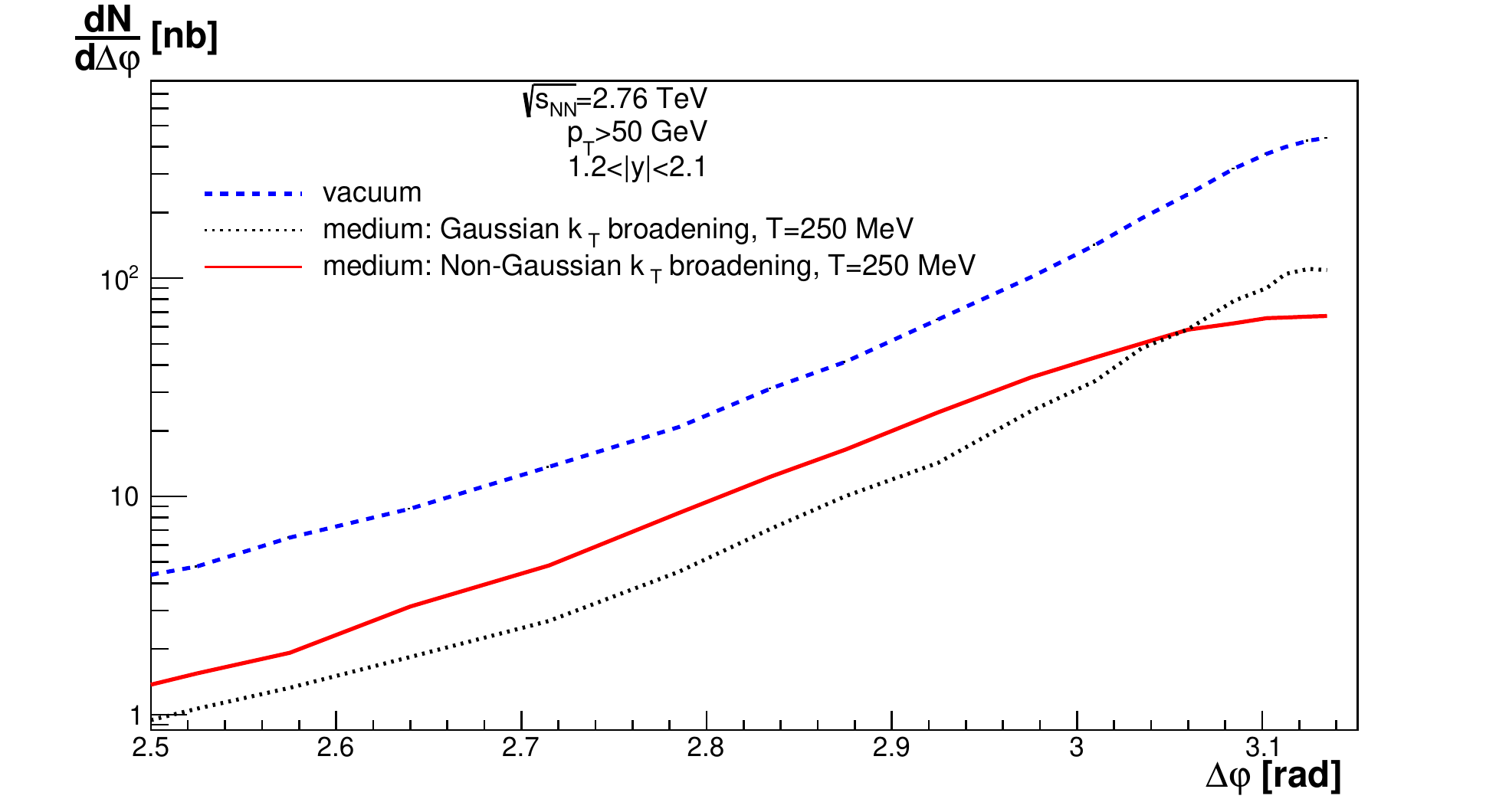}
\includegraphics[scale=0.4,clip=true,trim=20 5 45 0]{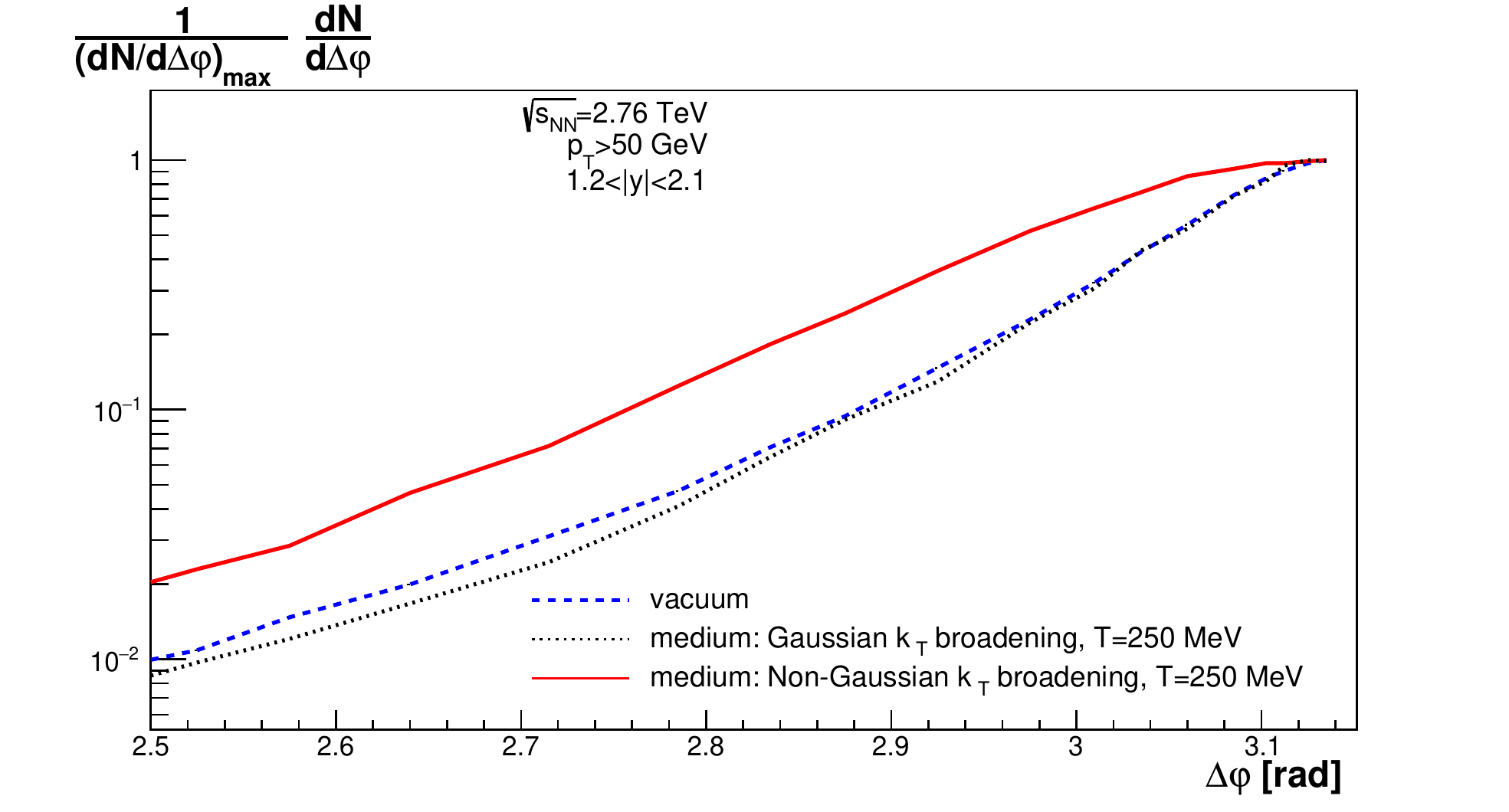}
\caption{Left panel: results for azimuthal angular decorrelations for di-jets produced in nuclear collisions at $\sqrt{s_{\rm NN}}=2.76$~TeV.
Right panel: same curves, but now normalized to their respective maxima.}
\label{fig1}
\end{figure}

For jet-pairs produced in collisions at $\sqrt{s_{\rm NN}}=2.76$~TeV, where jets are emitted in  directions with rapidities $1.2$$<$$|y|$$<$$2.1$, the left panel of Fig.~\ref{fig1} shows results for $\frac{dN}{d\Delta\phi}$. 
There, the cases with jet-medium interactions are considerably suppressed. 
Differences in shapes of the cases are examined in the right panel of Fig.~\ref{fig1}, which shows the $\frac{dN}{d\Delta\phi}$ distributions for all three cases divided by their respective maxima.
The vacuum case behaves similarly to the case of Gaussian $k_T$ broadening, where the curves for the latter case correspond to an even slightly narrower distribution than that of the former.
However, the distribution for non-Gaussian $k_T$ broadening is considerably broader.

We studied di-jet production in heavy ion collisions by using a Monte-Carlo approach which accounts for the momentum components of the colliding partons transverse to the beam axis via unintegrated parton densities and  an in-medium jet evolution that follows Eqs.~(\ref{eq:ktee1}) and~(\ref{eq:qee1}) from~\cite{Blaizot:2014rla}  with both scattering off-medium particles and coherent medium induced radiation. 
Some shortcomings of our approach are that it does not include quark jets, that the leading jet-gluon momenta are taken as the jet momenta, and that we do not include bremsstrahlung emission in vacuum.
Thus, we do not use our approach for comparison with the experiment, but rather to qualitatively study the influences of in-medium interactions on di-jet observables.

We obtained phenomenological results for the azimuthal angular decorrelations $\frac{dN}{d\Delta\phi}$. 
We conclude that the obtained distribution for di-jets in a QGP-medium is considerably suppressed as compared to di-jet production without consideration of a medium and also considerably broader. The main reason of the broadening are deviations of the $k_T$ distributions from Gaussian behavior.
%
%
%
%
 %
\section*{Acknowledgements}
The research is supported by 
Polish National Science Centre grant no. DEC-2017/27/B/ST2/01985
\bibliographystyle{JHEP}
\bibliography{refs}

\providecommand{\href}[2]{#2}\begingroup\raggedright\begin{thebibliography}{1}

\bibitem{vanHameren:2016kkz}
A.~van Hameren, \emph{{KaTie : For parton-level event generation with
  $k_T$-dependent initial states}},
  \href{https://doi.org/10.1016/j.cpc.2017.11.005}{\emph{Comput. Phys. Commun.}
  {\bfseries 224} (2018) 371}
  [\href{https://arxiv.org/abs/1611.00680}{{\ttfamily 1611.00680}}].

\bibitem{Kutak:2018dim}
K.~Kutak, W.~Płaczek and R.~Straka, \emph{{Solutions of evolution equations
  for medium-induced QCD cascades}},
  \href{https://doi.org/10.1140/epjc/s10052-019-6838-9}{\emph{Eur. Phys. J.}
  {\bfseries C79} (2019) 317}
  [\href{https://arxiv.org/abs/1811.06390}{{\ttfamily 1811.06390}}].

\bibitem{Blaizot:2012fh}
J.-P.~Blaizot, F.~Dominguez, E.~Iancu and Y.~Mehtar-Tani, \emph{{Medium-induced
  gluon branching}}, \href{https://doi.org/10.1007/JHEP01(2013)143}{\emph{JHEP}
  {\bfseries 01} (2013) 143} [\href{https://arxiv.org/abs/1209.4585}{{\ttfamily
  1209.4585}}].

\bibitem{Blaizot:2013vha}
J.-P.~Blaizot, F.~Dominguez, E.~Iancu and Y.~Mehtar-Tani, \emph{{Probabilistic
  picture for medium-induced jet evolution}},
  \href{https://doi.org/10.1007/JHEP06(2014)075}{\emph{JHEP} {\bfseries 06}
  (2014) 075} [\href{https://arxiv.org/abs/1311.5823}{{\ttfamily 1311.5823}}].

\bibitem{Blaizot:2014rla}
J.-P.~Blaizot, L.~Fister and Y.~Mehtar-Tani, \emph{{Angular distribution of
  medium-induced QCD cascades}},
  \href{https://doi.org/10.1016/j.nuclphysa.2015.03.014}{\emph{Nucl. Phys.}
  {\bfseries A940} (2015) 67}
  [\href{https://arxiv.org/abs/1409.6202}{{\ttfamily 1409.6202}}].

\bibitem{Aurenche:2002pd}
P.~Aurenche, F.~Gelis and H.~Zaraket, \emph{{A Simple sum rule for the thermal
  gluon spectral function and applications}},
  \href{https://doi.org/10.1088/1126-6708/2002/05/043}{\emph{JHEP} {\bfseries
  05} (2002) 043} [\href{https://arxiv.org/abs/hep-ph/0204146}{{\ttfamily
  hep-ph/0204146}}].

\bibitem{vanHameren:2019xtr}
A.~van Hameren, K.~Kutak, W.~P\l{}aczek, M.~Rohrmoser and K.~Tywoniuk,
  \emph{{Jet quenching and effects of non-Gaussian transverse-momentum
  broadening on di-jet observables}},
  \href{https://arxiv.org/abs/1911.05463}{{\ttfamily 1911.05463}}.

\end{thebibliography}\endgroup

\end{document}